

\magnification=\magstep1
\baselineskip=18 pt
%
%
\def\wc{\hangindent=4em \hangafter=1 \noindent}
\def\Ref #1 {\lbrack {#1}\rbrack}
\def\ref#1  {\noindent \hangindent=24.0pt \hangafter=1 {#1} \par}
\def\vol#1  {{{#1}{\rm,}\ }}
\def\etal   {{et~al.}\ }
\def\apj    {{ApJ{\rm,}\ }}

\def\aa     {{A\&A{\rm,}\ }}

\def\anrev  {{ARA\&A{\rm,}\ }}

\def\mnras  {{MNRAS{\rm,}\ }}

\def\pasp   {{PASP{\rm,}\ }}
\def\nod {\vskip 0.183 cm \noindent}
\def\section#1{\medbreak \centerline{\bf #1} \medskip}
\def\b {$\beta$ }
\def\bff {$\beta_{fit}$ }
\def\bfc {$\beta_{fit}^c$ }
\def\bs {$\beta_{spec}$ }
%
%
\def\today{\ifcase\month\or
  January\or February\or March\or April\or May\or June\or
  July\or August\or September\or October\or November\or December\fi
  \space\number\day, \number\year}
%
%
%
%
%
%
%
\voffset 3cm
\centerline{\bf RESOLVING THE BETA -- DISCREPANCY FOR CLUSTERS OF GALAXIES}

\vskip 1.0cm
\centerline {Neta A. Bahcall \& Lori M. Lubin}
\vskip 0.5cm
\centerline {\it Princeton University Observatory}
\centerline {\it Peyton Hall, Princeton, NJ 08544-1001, U.S.A.}
\vskip 6cm
\centerline {To be published in {\it The Astrophysical Journal}}
%
\vfill
\eject
\voffset 0 cm
\vskip 2 cm

\section{ABSTRACT}
Previous comparisons of optical and X-ray observations of
clusters of galaxies have led to the so-called ``$\beta$ --
discrepancy'' that has persisted for the last decade.  The standard
hydrostatic-isothermal model for clusters predicts that the parameter
$\beta_{spec} \equiv \sigma_{r}^{2}/(kT/\mu m_{p})$, which describes the
ratio of energy per unit mass in galaxies to that in the gas, should
equal the parameter $\beta_{fit}$ (where $\rho_{gas}(r) \propto
\rho_{gal}(r)^{\beta_{fit}}$) determined from the X-ray surface
brightness distribution.  The observations suggest an apparent
discrepancy : $\beta_{spec} \sim 1.2$ (i.e., the galaxies are
``hotter'' that the gas) while $\beta_{fit} \sim 0.65$ (i.e., the gas
is ``hotter'' and more extended than the galaxies).  Here we show that
the discrepancy is resolved when the actual observed galaxy
distribution in clusters is used, $\rho_{gal}(r) \propto r^{-2.4 \pm
0.2}$, instead of the previously assumed steeper King approximation,
$\rho_{gal}(r) \propto r^{-3}$.  Using the correct galaxy profile in
clusters, we show that the standard hydrostatic-isothermal model
predicts $\beta_{spec} = \beta_{fit}^{c} \simeq (1.25 \pm 0.10) \times
\beta_{fit}$, rather than $\beta_{spec} \simeq \beta_{fit}$ (where
$\beta_{fit}$ is the standard parameter using the King approximation,
and $\beta_{fit}^{c}$ is the corrected parameter using the proper
galaxy distribution).  Using a large sample of clusters, we find
best-fit mean values of $\beta_{spec} = 0.94 \pm 0.08$ and
$\beta_{fit}^{c} = 1.25 \times \beta_{fit} = 0.84 \pm 0.10$.  These
results resolve the $\beta$ -- discrepancy and provide additional
support for the hydrostatic cluster model.

\vskip 1 cm

\wc{{\it Subject headings}: Galaxies: clustering -- X-rays: galaxies \hfill}

\vfill
\eject

\section{1. INTRODUCTION}

The gravitaional mass of clusters of galaxies has traditionally been
estimated from the dynamics of the cluster galaxies using the virial
theorem, revealing large amounts of dark matter (Zwicky 1933, Peebles 1980).
More recently, X-ray emission from the hot intracluster gas has been used
to estimate cluster masses by utilizing the intracluster gas temperature
and the density profile as a tracer of the cluster potential assuming
hydrostatic equilibrium for the cluster gas (Cavaliere \& Fusco-Femiano
1976, Bahcall \& Sarazin 1977, Forman \& Jones 1984, Sarazin 1986,
Hughes 1989, Evrard 1990, Bahcall \& Cen 1993).  The two methods yield
comparable masses.  However, one problem has persisted over the years :
the so-called ``\b-- discrepancy'' of clusters.  This problem reflects the
discrepancy between the observed parameter \bs, determined from the X-ray
temperature and cluster velocity dispersion, and \bff, determined from the
gas versus galaxy density profiles in the clusters.  If the standard
hydrostatic cluster model is correct, the two parameters are expected to have
similar values; however, observations suggest that \bs $\sim (1.5 - 2) \times$
\bff (Sarazin 1986, Evrard 1990).  This discrepancy has been an unsolved
puzzle in the study of clusters of galaxies for nearly a decade.

In this paper, we show that the ``\b-- discrepancy'' results mainly from
assuming a galaxy distribution in clusters which is too steep (the King
approximation), thus causing misleading conclusions.  The ``\b-- discrepancy''
is resolved if the actual observed profile of the galaxy distribution in
clusters is used.

\section{2. THE \b-- DISCREPANCY}

The standard model for the structure of mass in clusters assumes that both
the gas and the galaxies are in hydrostatic equilibrium with the binding
cluster potential (Cavaliere \& Fusco-Femiano 1976, Bahcall \& Sarazin 1977,
Forman \& Jones 1984, Sarazin 1986, Evrard 1990).  In this model the gas
distribution obeys ${1\over{\rho_{gas}}}~{{dP_{gas}}\over{dr}} =
{{d\phi}\over{dr}} = - {{G M(r)}\over{r^2}}$, where $P_{gas}$ and $\rho_{gas}$
are the gas pressure
and density profiles, $\phi$ is the cluster potential, and $M(r)$ is the
total binding cluster mass within radius $r$ of the cluster center.  The
cluster mass can thus be represented as
$$
M(r) = - {{k T}\over{\mu m_p G}} \Bigl({{dln \rho_{gas}(r)}\over{dln r}} +
{{dln T}\over {dln r}}\bigr)~r	\eqno(1)
$$
where $T$ is the intracluster gas temperature, and $\mu m_p$ is the mean
particle mass of the gas.  The galaxies in the cluster respond to the same
gravitational field, and thus satisfy
$$
M(r) = - {{\sigma_r^2}\over{G}} \Bigl({{dln \rho_{gal}(r)}\over{dln r}} +
{{dln \sigma_r^2}\over {dln r}} + 2 A\bigr)~r	\eqno(2)
$$
where $\sigma_r$ is the radial velocity dispersion of the galaxies in the
cluster, $\rho_{gal}(r)$ is the galaxy density as a function of $r$, and
$A$ represents a possible anisotropy in the galaxy velocity distribution
[$A = 1 - (\sigma_t/\sigma_r)^2$, where $t$ and $r$ represent the tangential
and radial velocity components].

Relations (1) and (2) thus yield :
$$
\beta_{spec} \equiv {{\sigma_r^2}\over{k T/\mu m_p}} =
{{d ln \rho_{gas}(r)/d ln r + d ln T/d ln r}\over{d ln \rho_{gal}(r)/d ln r +
d ln \sigma_r^2/d ln r + 2 A}}	\eqno(3)
$$
where the \bs parameter is defined by the left side of relation (3).  This
spectral \b parameter can be determined directly from observations of
cluster velocity dispersions and gas temperatures; it represents the ratio
of energy per unit mass in the galaxies to that in the gas and is observed
to be $\sim 1$ (see below).  (Here and below we assume that the cluster
velocity dispersion is isotropic, thus the radial velocity dispersion
$\sigma_r$ is comparable to the observed line-of-sight velocity dispersion
$\sigma$).  The right-hand side of equation (3) relates the \bs parameter to
the density profiles of the gas and galaxies in the cluster as well as to
the temperature and velocity profiles.

For isothermal galaxy and gas distributions and isotropic galaxy velocities
(a model that is traditionally used to describe clusters), one has $d ln T/ d
ln r = d ln \sigma_r^2/d ln r = A \simeq 0$; relation (3) then becomes
$$
\beta_{spec} \equiv {{\sigma_r^2}\over{k T/\mu m_p}} =
{{d ln \rho_{gas}(r)/d ln r}\over{d ln \rho_{gal}(r)/d ln r}} \equiv
\beta_{fit}^c	\eqno(4)
$$
the \bs parameter determined from $\sigma_r$ and $T$ (left side of
eq. 4) should approximately be equal to the ratio of the slopes of the
gas to galaxy density profile, defined as \bfc (right side of equation 4).
The solution of equation (4) yields $\rho_{gas}(r) \propto
\rho_{gal}(r)^{\beta_{fit}^c}$ and $\beta_{spec} \simeq \beta_{fit}^c$
. The parameter \bfc is determined
by fitting the observed X-ray surface brightness distribution in clusters.
The galaxy distribution needed to normalize the \bfc determination has been
generally {\it assumed} to be the King (1962) approximation, $\rho_{gal}(r)
= \rho_{gal}(0) {(1 + (r/R_c)^2)}^{-3/2}$, where $\rho_{gal}(r)
\propto r^{-3}$ for large radii (i.e., $r > R_c$, where $R_c$ is the cluster
core radius). Therefore, $\rho_{gas}(r) = \rho_{gas}(0) {(1 +
(r/R_c)^2)}^{-3 \beta_{fit}/2}$; here \bff
without the superscript ``c'' for ``correct'' denotes the standard definition
of \bff using the King approximation (see, e.g. Abramopoulos \& Ku 1983,
Jones \& Forman 1984, Sarazin 1986, Evrard 1990, and references therein).
Therefore, \bff has previously been determined from the relation
$$
\rho_{gas}(r)~\lbrack {\propto \rho_{gas}(r)^{\beta_{fit}^c}}\rbrack
\propto r^{-3 \beta_{fit}}	\eqno(5)
$$
where \bff represents the determination using the King approximation, and
\bfc is the ``correct'' value from equation (4).

X-ray observations (Abramopoulos \& Ku 1983, Jones \& Forman 1984, Henry
et al. 1993) yield \bff values in the range $\sim$ 0.5 to 0.9, with a median
of \bff $\simeq 0.67 \pm 0.10$ (rms); this corresponds to a gas density
profile of
$$
\rho_{gas}(r) \propto r^{-2.0 \pm 0.3} {\hskip 2cm}  R_c < r \le 1.5~h^{-1}~Mpc
\eqno(6)
$$
On the other hand, observations yield \bs $= \sigma_r^2/(k T/\mu m_p)$
values in the range of $\sim$
0.5 to $\sim$ 2 with a median \bs $\simeq 1 - 1.4$ (Mushotsky 1984, Sarazin
1986, Evrard 1990, Edge \& Stewart 1991, Lubin \& Bahcall 1993).  This
inequality between \bs and \bff contradicts the expectation from the
hydrostatic
model.  We discuss below how this discrepency is resolved.

\section{3. RESOLVING THE \b-- DISCREPANCY}

The ``\b-- discrepancy'' is based on the assumption that the galaxy
distribution in clusters follows the King approximation, $\rho_{gal}(r)
\propto r^{-3}$ at large radii.  This assumption, however, is inaccurate.
Specific studies of the galaxy distribution in clusters yield a shallower
profile.  The galaxy-cluster cross-correlation function, which represents
the {\it average} net galaxy density profile around rich clusters, yields
$\rho_{gal}(r) \propto r^{-2.2}$ (Lilje \& Efstathiou 1988), or $\rho_{gal}(r)
\propto r^{-2.4}$ (Seldner \& Peebles 1978, Peebles 1980).  The average
profile of a sample of rich clusters yields $\rho_{gal}(r) \propto r^{-2.6}$
(Bahcall 1977) (all for $0.5~h^{-1} \le r \le 1.5~h^{-1}~Mpc$). The profile
becomes even shallower, $\rho_{gal}(r) \propto r^{-2}$ for $r \le
0.5~h^{-1}~Mpc$ (above references; see also Beers \& Tonry 1986).  We use
below the observed range of the average galaxy density distribution in
rich clusters
$$
\rho_{gal}(r) \propto r^{-2.4 \pm 0.2} {\hskip 2cm} 0.5~h^{-1} < r \le
1.5~h^{-1}~Mpc	\eqno (7)
$$
Inserting relations (6--7) in equations (4--5) we find :
$$
\rho_{gas}(r) \propto \rho_{gal}(r)^{\beta_{fit}^c} \propto r^{-(2.4 \pm 0.2)
\beta_{fit}^c} \propto r^{-3 \beta_{fit}} \propto r^{-2.0 \pm 0.3}
\eqno(8)
$$
where the last term in (8) represents the observed X-ray profiles (Sect. 2).
The corrected \bfc parameter introduced here is related to the parameter
determined using the King approximation, \bff (eq. 5), by
$$
\beta_{fit}^c \simeq {{3}\over{2.4 \pm 0.2}}~\beta_{fit} = (1.25 \pm 0.10)
\times \beta_{fit}	\eqno(9)
$$
Since the X-ray observations yield a mean value of \bff $= 0.67\pm 0.02$
(Jones \& Forman 1984, Henry et al. 1993), the corrected \bfc parameter
(eq. 9) has a mean value of
$$
\beta_{fit}^c = (1.25 \pm 0.10) \times \beta_{fit} = 0.84 \pm 0.10
\eqno(10)
$$
If the fit is dominated by X-ray emission at small separations, where
$\rho_{gal} \propto r^{-2}$, then \bfc $\sim 1$.

Detailed X-ray and optical observations of the cluster A2256 (Henry et al.
1993) using recent X-ray data from ROSAT support the above analysis; the
observations yield $\rho_{gas} \propto r^{-2.4}$, $\rho_{gal} \propto
r^{-2.5}$, \bff $= 0.795$, and therefore \bfc $=0.96$, fully consistent
with relations (7--10).

A recent analysis of the largest-available sample of clusters with measured
$T$ and line-of-sight velocity dispersions $\sigma$ (each having more than
twenty galaxy redshifts per cluster) yields a best-fit \bs parameter for the
entire sample (Lubin \& Bahcall 1993) :
$$
\beta_{spec} \equiv {{\sigma_r^2}\over{k T/\mu m_p}} \approx
{{\sigma^2}\over{k T/\mu m_p}} = 0.94 \pm 0.08
\eqno(11)
$$
This best-fit \bs was obtained from a $\chi^2$ fit to the observed $\sigma(T)$
relation for the sample [$\sigma = (\beta_{spec}/\mu m_p)^{0.5}~(k T)^{0.5}$]
using the average observed $\mu = 0.58$ for
clusters (Edge \& Stewart 1991).  The median value for the sample is
\bs$(median) = 0.98$.

The directly observed values of \bs and \bfc (eqs. 10--11) are consistent
with each other as expected from the hydrostatic-isothermal model (eq. 4).
Various effects such as possible substructure in some clusters (Geller \&
Beers 1982), velocity anisotropy or contamination, cooling flows, or
incomplete thermalization of the gas may contribute to the observed scatter
(see Figures below) but do not significantly affect the main results obtained
above (see Lubin \& Bahcall 1993).

The agreement between \bs and \bfc persists even if the distributions are
not isothermal.  Some clusters show a decrease in their temperature and
velocity dispersion profiles (e.g., Coma; Kent \& Gunn 1982, Jones \&
Forman 1992).  Using relation (3), with temperature and velocity gradients of
$d ln T/d ln r \simeq d ln \sigma_r^2/d ln r \sim -0.5$ to --1 (as appears to
be suggested by the results in the above references), as well
as allowing for a small amount of velocity anistropy in clusters ($A \le 0.2$),
we find \bfc $\sim 0.9 - 1$ (from the right side of eq. 3).  This value is
in excellent agreement with the directly measured value of \bs $= 0.94 \pm
0.08$ (eq. 11).

The results obtained here are illustrated in Figures 1--3.  In Figure 1, we
plot the observed \bfc and \bs values for all clusters where both parameters
have been measured (Jones \& Forman 1984 for \bff; Lubin and Bahcall 1993 for
\bs).  The original \bff values (assuming the King approximation) are
presented in Figure 1a, while the corrected \bfc values are presented in
Figure 1b.  While Figure 1a clearly reflects the \b-- discrepancy (i.e.
$\beta_{spec} > \beta_{fit}$), this discrepancy is eliminated in Figure 1b,
where the corrected \bfc values (corresponding to a galaxy density profile
of $\rho_{gal} \propto r^{-2.4 \pm 0.2}$) are plotted.  The ratio \bs/\bfc
is presented for all clusters in Figure 2.  It is clear from the results
in Figure 2 that, within the observational uncertainties, no \b-- discrepancy
is apparent (\bs/\bfc $\sim 1$).  Finally, we plot in Figure 3 the recently
observed gas and galaxy density profiles in the cluster A2256 (Henry et al.
1993).  The data supports the main conclusions reached above : the gas and
galaxy distributions follow each other (i.e., \bfc $\simeq 1$) within the
observational uncertainties.  Both distributions can be represented by
an $\rho \propto r^{-2.5}$ profile.  The spectroscopic \bs parameter of
the cluster is \bs $=1.33 \pm 0.24$ (Lubin \& Bahcall 1993); these values are
included in Figures 1--2 and are consistent with the entire set presented.

We conclude that no significant \b-- discrepancy exists in clusters of
galaxies.  The hydrostatic model therefore provides a consistent fit to both
the X-ray and the optical data.

\vskip 0.5 cm
We would like to thank J.P. Ostriker, P.J.E. Peebles, D. Richstone, C.L.
Sarazin, and D.N. Spergel for helpful discussions.
\vfill
\eject

\section{FIGURE CAPTIONS}
\noindent
{\bf Fig. 1}
\nod
(a) The observed values of the parameters \bff (using the King approximation;
Jones \& Forman 1984) and \bs ($= \sigma^2/k T/\mu m_p$; Lubin \& Bahcall
1993) for all clusters for which both parameters are measured.  The
``\b-- discrepancy'' effect, i.e., $\beta_{spec} > \beta_{fit}$, is apparent

\nod
(b) Same as (a), but for the corrected \bfc values, assuming the average
observed galaxy distribution $\rho_{gal}(r) \propto r^{-2.4 \pm 0.2}$
(Sect. 3).  No ``\b-- discrepancy'' is apparent. (The scatter in \bs is larger
than in \bfc, as expected due to larger observational uncertainties.)
\nod
{\bf Fig. 2}
\nod
The ratio \bs/\bfc for all avaliable clusters (same data as Figure 1b).  No
apparent discrepancy is seen (i.e., \bs/\bfc $\sim 1$).  Different symbols
represent different sub-samples : filled squares -- clusters in superclusters;
triangles -- isolated clusters; open squares -- clusters at $|b| < 20^{o}$
(see Lubin \& Bahcall 1993).
\nod
{\bf Fig. 3}
\nod
Projected galaxy density distribution (open circles) and projected gas density
distribution (solid curves represent $\pm 1$ sigma results) for the cluster
A2256 (Henry et al. 1993).  The galaxy density is in units of galaxies per
deg$^2$; the gas density is in arbitrary units.  Both profiles follow
$\rho(r) \propto r^{-2.5}$ outside the cluster core.
\vfill
\eject

\section{REFERENCES}
\vskip 0.3cm
\normalbaselineskip=8pt plus0pt minus0pt
                            \parskip 0pt
\ref{Abramopoulos, F., \& Ku, W. 1983, \apj\vol{271} 446}
\ref{Bahcall, N.A. 1977, \anrev\vol{15} 505}
\ref{Bahcall, N.A., \& Cen, R.Y. 1993, \apj\vol{407} L49}
\ref{Bahcall, N.A., \& Sarazin, C.L. 1977, \apj\vol{213} L99}
\ref{Beers, T., \& Tonry, J. 1986, \apj\vol{300} 557}
\ref{Cavaliere, A., \& Fusco-Femiano, R. 1976, \aa\vol{49} 137}
\ref{Edge, A.C., \& Stewart, G.C. 1991, \mnras\vol{252} 428}
\ref{Evrard, A.E. 1990, Clusters of Galaxies, eds. W.R. Oegerle \etal,
STScI Symposium 4, 287}
\ref{Geller, M.J., \& Beers, T. 1982, \pasp\vol{94} 421}
\ref{Henry, J.P., Briel, U.G., \& Nulsen, P.E. 1993, \aa\vol{271} 413}
\ref{Hughes, J.P. 1989, \apj\vol{337} 21}
\ref{Jones, C., \& Forman, W. 1984, \apj\vol{276} 38}
\ref{Jones, C., \& Forman, W. 1992, Clusters and Superclusters of Galaxies,
ed. A.C. Fabian, NATO ASI Series 366, Kluwer Academic Publishers}
\ref{Kent, S.M., \& Gunn, J.E. 1982, \apj\vol{87} 945}
\ref{King, I. 1972, \apj\vol{174} L123}
\ref{Lilje, P.B., \& Efstathiou, G. 1988, \mnras\vol{231} 635}
\ref{Lubin, L.M. \& Bahcall, N.A. 1993, \apj\vol{415} L17}
\ref{Mushotzky, R. 1984, Phys. Scripta, \vol{T7} L157}
\ref{Peebles, P.J.E. 1980, The Large Scale Structure of the Universe,
Princeton University Press, Princeton}
\ref{Sarazin, C.L. 1986, Rev. Mod. Phys., \vol{58} 1}
\ref{Seldner, M., Peebles, P.J.E. 1977, \apj\vol{215} 703}
\ref{Zwicky, F. 133, Helv. Phys. Acta, 6, 110}
\vfill
\eject
\end